\documentstyle[pre,twocolumn,graphicx,psfig,amsfonts,aps]{revtex}
\begin{document}

\newcommand{\beqa}{\begin{eqnarray}}
\newcommand{\eeqa}{\end{eqnarray}}
\newcommand{\beq}{\begin{equation}}
\newcommand{\eeq}{\end{equation}}
\newcommand{\bfit}[1]{\mbox{\boldmath $#1$}}

\draft
\title{On two continuous models for the dynamics of sandpile surfaces}
\author{Leonid Prigozhin\thanks{E-mail: leonid@bgumail.bgu.ac.il}
and Boris Zaltzman\thanks{E-mail: boris@bgumail.bgu.ac.il}} 
\address{Department for Solar Energy and Environmental Physics\\
Blaustein Institute
for Desert Research\\
Ben Gurion University of the Negev, Sede Boqer 
Campus, 84990 Israel}
\date{May 15, 2000}
\maketitle
\begin{abstract}
We consider a modified BCRE model for pile surface dynamics and show that in the 
long-scale limit this model converges to a quasistationary model of pile growth
in the form of an evolutionary variational inequality. 
\end{abstract} 

\section{Introduction}

Recent much interest to the physics of granular media was, in particular,
stimulated by two salient features of the granular state: multiplicity
of metastable pile shapes and occurrence of avalanches upon pile surfaces.
It has been realized that, to account for metastability, the model of pile
surface dynamics should not be written as an evolutionary equation for the
pile surface alone. An additional unknown characterizing the flow of grains
down the pile surface is useful because such flows are not uniquely
determined by the external source and local free surface topography.

A large spatio-temporal scale pile growth model involving two coupled
dependent variables and able to account for metastability has been proposed
in \cite{ZhVM,PRE1}. This model neglects avalanches as small fluctuations of
the pile surface and describes the evolving mean surface of a pile that
grows on an arbitrary support under a given distributed source of bulk
material. The model permits an equivalent formulation as an evolutionary
variational or quasivariational inequality; such a formulation simplifies
significantly both the mathematical study of the problem \cite{Euro1} and
its numerical solution \cite{ZhVM,ChEngSci,EV}. As it has been shown in \cite
{PRE1}, the shapes of real piles on flat open platforms \cite{Puhl} are
described by the analytical solutions of this inequality. A modification of
the model, able to account for avalanches as almost instantaneous slides, is
also discussed in \cite{PRE1}; according to observations made in the same work,
such a slide may, indeed, be a possible avalanche scenario (see also \cite
{Smith}).

Independently and using different arguments, the same pile growth model in
the form of a variational inequality has been derived by Aronsson, Evans,
and Wu \cite{Evans1}. In \cite{Evans2}, Evans {\it et al.} studied its
discontinuous solutions corresponding to avalanches; in \cite{Evans3}
Evans and Rezakhanlou showed that the cellular automata models of sandpiles,
presented as intuitively attractive examples in almost all works on
self-organized criticality \cite{Bak}, converge in a continuous limit to a
similar variational inequality with an anisotropy inherited from the
cellular structure of these crude models.

A different continuous model, also involving two coupled dependent variables
and describing the granular surface flow and pile surface dynamics, has been
proposed by Bouchaud, Cates, Ravi Prakash, and Edwards (the BCRE model) \cite
{BCRE1,BCRE2}. Although the choice of the basic variables in this model is
equivalent to that in \cite{ZhVM,PRE1}, the model is written for the free
surfaces only slightly deviating from the critical slope and employs
different phenomenological constitutive relations. The emphasis is put onto
the simulation of fast processes, like amplification and distinction of
rolling grains population during an avalanche. The BCRE model has been
simplified by de Gennes \cite{deG1}, applied to various one-dimensional
surface flow problems (see, e.g., \cite{deG2,deGA}), and modified for thick
surface granular flows \cite{Thick}. Further exact solutions to simplified
BCRE equations can be constructed by the methods proposed in \cite{MP}.
Using the BCRE model, Bouchaud and Cates \cite{BC} explained another type of
avalanches (in a thin granular layer on an inclined plane, see \cite{DD}).

Our aim here is to investigate a relation between the two models mentioned
above. After reminding briefly the variational and BCRE models, we propose a
full-dimensional generalization of the latter, originally formulated by BCRE
in the one-dimensional case. To do this, we modify and extend the constitutive
relations determining the surface flow velocity and the
rolling-to-immobilized-state transition rate: BCRE's assumption that the slope
is everywhere almost critical is too restrictive for our purpose. Rescaling
the variables, we show that the modified BCRE model contains a small
parameter, the ratio of a characteristic rolling grains layer thickness to
the pile size, and hence may often be simplified by employing a
quasistationary equation for the rolling grains layer. The issue of scaling
turns out to be very important in description of pile growth: another
dimensionless parameter in the model thus obtained is the ratio of a typical
rolling grain path length to the pile size. For large piles, this
coefficient is also small and we show that in the long-scale limit the
modified BCRE model tends to the variational model \cite{ZhVM,PRE1}. For
small piles, the corresponding term can be significant. These results make clear 
why differ the shapes of small and large piles and,
correspondingly, why different models should be used to simulate, say,
formation of large sand dunes and small Aeolian ripples.

\section{Variational model of pile growth}

Let a cohesionless granular material having an angle of repose $\alpha_r$ be
tipped out onto a given rough rigid surface $y=h_0(x)$,
where $x=(x_1,x_2)\in \mathbb{R}$$^2$. We want to find the shape of a pile
thus generated.

The real process of pile growth is often intermittent: discharged granular
material not only flows continuously over the pile slopes but is also able
to build up and then to pour suddenly down the slope in an avalanche.
However, the avalanches usually involve only a small amount of particles in
a pile and cause small fluctuations of the pile free surface. The model \cite
{ZhVM,PRE1} neglects these fluctuations and is a model for the mean surface
evolution. Whether the pile evolution is governed by a continuous surface
flow or results from many small avalanches, the surface flow is typically
confined to a thin boundary layer which is distinctly separated from the
motionless bulk \cite{JN}.

Let us assume for simplicity that the support surface has no steep slopes,
i.e., 
\[
|\nabla h_0|\le k,
\]
where $k=\tan\alpha_r$ (see \cite{ZhVM,PRE1,Euro1} for the general case).
Assuming the bulk density of material in a pile is constant we can write the
conservation law as 
\[
\partial_th+\nabla\cdot\mbox{\boldmath $q$}=w,
\]
where $h(x,t)$ is the free surface, $\mbox{\boldmath $q$}
(x,t)$ is the horizontal projection of the flux of rolling
particles, and $w(x,t)$ -- the source intensity. We
neglect the inertia and suppose that surface flow is directed towards the
steepest descent, 
\[
\mbox{\boldmath $q$}=-m\nabla h,
\]
where 
\begin{equation}
m(x,t)\ge 0  \label{m}
\end{equation}
is an {\it unknown} scalar function. The conservation law takes now the form 
\begin{equation}
\partial_th-\nabla\cdot(m\nabla h)=w.  \label{conserv}
\end{equation}
It is assumed in this model that the surface slope angle cannot exceed the
angle of repose, 
\begin{equation}
|\nabla h|\le k,  \label{cond1}
\end{equation}
and that no pouring occurs over the parts of the pile surface which are
inclined less: 
\begin{equation}
|\nabla h(x,t)|< k\ \Longrightarrow\ m(x,t)=0.  \label{cond2}
\end{equation}
To complete the model we have to specify the initial, 
\begin{equation}
h|_{t=0}=h_0,\label{t0}
\end{equation}
and a boundary condition. Let the granular material be allowed to leave the
system freely through part $\Gamma_1$ of the boundary of domain $%
\Omega\subset \mathbb{R}$$^2$, and the other part of the boundary, $\Gamma_2$, presents
an impermeable wall. The boundary conditions are then, respectively, 
\begin{equation}
h|_{\Gamma_1}=h_0|_{\Gamma_1},\ \ \ \ m\partial_{n}h|_{\Gamma_2}=0.
\label{bound}
\end{equation}

\begin{figure}[hbt]
\centerline{\includegraphics[height=2.5cm]{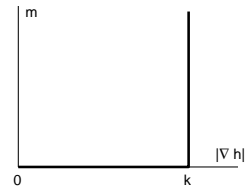}}
\caption{Multivalued constitutive relation $m=m(|\nabla h|)$.}
\label{graph1}
\end{figure}

The model (\ref{m})-(\ref{bound}) contains two coupled unknowns, the free
surface $h$ and an auxiliary function $m$ determining the rolling grains
flux magnitude. Conditions (\ref{m}), (\ref{cond1}), and (\ref{cond2})
define $m$ as a {\it multivalued} function of $|\nabla h|$, see Fig.~\ref
{graph1}. 
The problem (\ref{m})-(\ref{bound}) may be considered an anomalous
diffusion problem and solved by approximating this highly nonlinear
multivalued relation. However, a better way to solve this problem is based
on its following reformulation in the form of an evolutionary variational inequality
(see \cite{DL} and \cite{GLT} for variational inequalities in mechanics and
physics and their numerical solution, respectively).

Let us define the set $K$ of possible surfaces as 
\[
K=\left\{ \varphi (x)\ \ \bracevert \ \ |\nabla \varphi |\leq k,\ \
\varphi |_{\Gamma _{1}}=h_{0}|_{\Gamma _{1}}\right\} 
\]
and the scalar product of two functions as $(\phi ,\psi )=\int_{\Omega }\phi
\psi \,dx$. We can now consider the following problem
(variational inequality): 
\begin{equation}
\left\{ 
\begin{array}{c}
Find\text{ \ \ }h(x,t)\text{ \ }such\text{ }that\text{ }%
h\in K\text{ }for\text{ }all\text{ }t>0, \\ 
(\partial _{t}h-w,\varphi -h)\geq 0\ \ \mbox{\it for all}\ \varphi \in K, \\ 
\mbox{\it and}\ h|_{t=0}=h_{0}.
\end{array}
\right.   \label{vi}
\end{equation}

Theorem. {\it Function $h(x,t)$ is a solution of the
variational inequality} (\ref{vi}) {\it if and only if there exists $m(x,t)$ 
such that the pair $\{h,m\}$ is a solution to} (\ref
{m})-(\ref{bound}).

The outline of the proof is given in \cite{PRE1} (see \cite{Euro1} for
mathematical details and a proof of existence of a unique solution to the
variational inequality (\ref{vi})). It has been also shown that the surface
flux magnitude $m(x,t)$ is, in this model, a Lagrange multiplier
related to the point-wise constraint (\ref{cond1}). 
The values of such multipliers are not uniquely determined by the local
conditions, which is the "mathematical explanation" of long-range
interactions typical of extended dissipative systems in a critical
(marginally stable) state, see \cite{FBP}.

The model (\ref{m})-(\ref{bound}) or, equivalently, (\ref{vi}), has simple
analytical solutions \cite{PRE1}, such as the conical pile growing under a
point-like source or the piles on flat open platforms described in \cite
{Puhl}. Numerical solutions \cite{ZhVM,ChEngSci} also demonstrate simple
geometrical structures that agree with one's sandbox memories. Although
this model is much simplified in many respects, it allows for 
the multiplicity of possible pile shapes. The avalanches may be introduced into 
the model as solution discontinuities (in time)
triggered by sudden changes of the admissible set $K$ \cite{Evans2,PRE1} 
and are instantaneous
events. On the
time scale of a slow pile growth 
the life of an avalanche is, indeed, very short.

\section{Modified BCRE equations}

The BCRE equations \cite{BCRE1,BCRE2} involve two coupled variables: the
pile height, $y=h(x,t)$, and the effective thickness (density) of the
rolling grains layer, $R(x,t)$ ($R(x,t)d\Omega$ is the volume that the
material, currently rolling above the area $d\Omega$, would occupy in the
pile). The model has been formulated for a two-dimensional pile ($x\in \mathbb{R}$$^1$); 
free surface slope deviations from the critical angle were assumed small.
Original BCRE equations included diffusion terms to account for a
non-locality of grains dislodgement and for fluctuations of rolling grains
velocity. Although diffusion plays a crucial role in BCRE's scenario of
avalanches \cite{BCRE1,BCRE2,BC}, these terms were regularly omitted by other 
researchers who
either assumed that in their problems diffusion is insignificant and 
simplified the model, or proposed a different avalanche scenario (see,
e.g., \cite{deG1,deGA,MP,DM}). Below, we also omit the diffusion terms at
first but introduce small diffusion at a later stage as a means for model
regularization in transition to a large-scale limit.

Simplified BCRE equations may be written as follows: 
\[
\partial_th=\Gamma[h,R],\ \ \ \partial_tR+\partial_x(vR)=w-\Gamma[h,R]. 
\]
Here the term $\Gamma[h,R]$ accounts for the conversion of rolling grains
into immobilized grains and vice versa, $v$ is the horizontal projection of
rolling grains velocity, and $w(x,t)$ is the source intensity (we
assume that the tipped grains do not stick to the pile surface but join the
rolling grains first).

Limiting their consideration to the slopes that are close to critical, BCRE
assumed constant downslope drift velocity $v$. The surface flux
magnitude, $q=vR$, is thus determined solely by the rolling layer
thickness $R$. Since in the previous model $q= mk$ for the critical slopes, $m$
and $R$ play similar roles and the two choices
of basic variables, $\{h,m\}$ and $\{h,R\}$, are essentially equivalent. The
exchange term $\Gamma$ in BCRE model is linearized in a vicinity of the
critical angle $\alpha_r$ and is proportional (for thin surface flows) to $R$%
: $\Gamma[h,R]=\gamma R(\alpha_r-\theta),$ where $\theta(x,t)$ is the
surface slope angle and $\gamma$ is a coefficient.

For a three-dimensional pile ($x\in \mathbb{R}$$^2$), the model
equations are similar, 
\begin{eqnarray}
\partial_th=\Gamma[h,R],\ \ \ \partial_tR+\nabla\cdot(\mbox{\boldmath $v$}%
R)=w-\Gamma[h,R],  \label{eqhR}
\end{eqnarray}
but the constitutive relations determining $\mbox{\boldmath $v$}$ and,
probably, $\Gamma[h,R]$ should be modified; here we will follow \cite{Ripple}
(see also \cite{HK}). 
We assume that the rolling particles drift towards the steepest descent 
of the free surface with a mean velocity $v$ depending on the slope angle
(the steeper the slope, the higher is the velocity). On their way downslope,
these particles may be trapped and absorbed into the motionless bulk (the
steeper the slope, the lower is the trapping rate $\Gamma$). If the surface
is horizontal, the mean flow velocity is zero and the trapping rate is
maximal, for $\theta = \alpha_r$ the rolling
particles follow without trapping. 
Below, we
will not consider the  overcritical slopes and assume also that the trapping
rate is proportional to the amount of rolling grains $R$.

At least partially, this simplified picture can be justified by recent
experimental, theoretical, and numerical studies on the motion of a
spherical particle on a rough inclined plane \cite{rough,AI}. For the relevant
region of slope angles $\theta$, the energy dissipation due to the
multiple shocks experienced by a moving particle is equivalent to the
action of a viscous friction force \cite{rough}. Because of that such
particles reach a constant mean velocity proportional to $\sin\theta$.
Sometimes, however, the particles are suddenly trapped in a well and
completely lose their momentum in the direction of motion \cite{AI}.
Of course, conditions in the
collective flow of grains over the pile surface are somewhat different. In
particular, the flow velocity may depend on the thickness of rolling grains
layer \cite{P} and the exchange rate is not exactly proportional to $R$ \cite
{Thick}. Various improved dependencies can be incorporated into the
model. The limiting behavior of the modified
BCRE model is, however, robust and does not depend on details. For clarity of 
presentation we will
consider the long-scale limit of a thin-flow model with the simplest
phenomenological relations determining the flow velocity and
rolling-to-immobilized-state transition.

Since the mean velocity of surface flow is proportional to $\sin \theta$ 
\cite{rough}, its horizontal projection $v$ is proportional to $\sin\theta
\cos \theta =\tan\theta/(1+\tan^2\theta)$. Postulating that the flow is in the steepest
descent direction, we obtain $\mbox{\boldmath 
$v$}=-\mu {\nabla h}/(1+|\nabla h|^2),$ where $\mu$ is a coefficient. Simplifying this
relation we assume 
\begin{equation}
\mbox{\boldmath $v$}=-\mu \nabla h.  \label{v}
\end{equation}
The exchange rate $\Gamma$ should not depend on the slope orientation and we
assume  it to be a smooth decreasing function of $|\nabla h|^2$ that becomes
zero for critical slopes. Assuming $\Gamma$ is proportional to $R$ (thin
flows) we arrive at 
\begin{equation}
\Gamma[h,R]=\gamma R\left(1-\frac{|\nabla h|^2}{k^2}\right)  \label{G}
\end{equation}
as the simplest constitutive relation \cite{Ripple}. 
We will now derive a dimensionless
formulation for the modified BCRE model (\ref{eqhR})-(\ref{G}).

The parameters in this model have the following dimensions: $[\gamma]={\cal T%
}^{-1}$ and $[\mu]={\cal LT}^{-1}$. Let us denote by $\overline{w}$ the
characteristic intensity of the external source; $[\overline{w}]={\cal LT}^{-1}$.
The three length scales characterizing the pile surface dynamics and
surface granular flow may be defined as follows:

\begin{itemize}
\item   typical thickness of the rolling grains layer, $L_{R}=\overline{w}%
/\gamma $;

\item  mean path of a rolling particle before it is trapped
strongly depends on the slope steepness but, for a fixed subcritical slope,
is proportional to the ratio $L_{P}=\mu /\gamma $ characterizing the
competition between rolling and trapping;

\item  the pile size $L$.
\end{itemize}

The time $T=L/\overline{w}$ needed for a source with  given intensity 
$\overline{w}$ to produce a pile of size $L$ may be used as a long time
scale. Rescaling the variables, 
\[
x^{\prime}=\frac{1}{L}x, \ h^{\prime}=%
\frac{1}{L}h, \ R^{\prime}=\frac{1}{L_R}R, \
w^{\prime}=\frac{1}{\overline{w}%
}w,\ t^{\prime}=\frac{1}{T}t,
\]
we arrive at the following dimensionless formulation: 
\begin{equation}
\partial_t h=\Gamma[h,R],  \label{hsc}
\end{equation}
\begin{equation}
\frac{L_R}{L}\partial_tR-\frac{L_P}{L}\nabla\cdot (R\nabla h)=w-\Gamma[h,R],
\label{Rsc}
\end{equation}
\begin{equation}
\Gamma[h,R]=R\left(1-\frac{|\nabla h|^2}{k^2}\right).  \label{Gsc}
\end{equation}
Typically, $L_R\ll L_P<L.$ The first coefficient in (\ref{Rsc}) is very
small, so it may often be possible to omit the corresponding term and use a
quasistationary equation for the rolling layer. Such an approach has already
been employed in simulation of the dynamics of sand ripples, see \cite
{Ripple}. The second coefficient, $L_P/L$, may be significant for small
piles, like sand ripples, but becomes small too for large piles. 
Further simplification of the model is then appropriate.

\section{The long-scale limit of BCRE model}

Let us denote $\nu=L_P/L$ and study the $\nu\rightarrow 0$ behavior of the
model (\ref{hsc})-(\ref{Gsc}). This limit corresponds to the case of large
piles ($L\gg L_P$). We want to show that in this limit the pile shape
evolution is described by the variational inequality (\ref{vi}) which
remains invariant under the rescaling employed.

Physically, the situation is clear: although the model (\ref{hsc})-(\ref{Gsc}) 
permits grains to roll down upon any inclined slope, the rolling particles
are quickly stopped and their paths are short comparing to the pile
size for all except the almost critical slopes. This is essentially what is
assumed in the model \cite{ZhVM,PRE1} which permits rolling upon the critical
slopes only. Mathematically, the situation is somewhat more complicated.

Since $L_{R}\ll L_{P}$, we assume $L_{R}/L$ is $o(\nu)$ and set 
$L_{R}/L=\nu\lambda(\nu),$ where $\lambda$ tends to zero as $\nu\rightarrow 0$.
Let us introduce a new variable, $m=\nu R$, define $\psi (u)=1-u^{2}/k^{2}$,
and rewrite the model (\ref{hsc})-(\ref{Gsc}) as
$$\partial_{t}h=\frac{m\psi (|\nabla h|)}{\nu},\ \ \lambda \partial_{t}m-\nabla \cdot
(m\nabla h)=w-\frac{m\psi (|\nabla h|)}{\nu }.$$ 
For any $\nu>0$ this system consists of two coupled hyperbolic equations.
The second equation, which can be regarded as an equation for $m$,
contains in its main part the coefficient $\nabla h$ which may be discontinuous.
The theory for such equations is complicated and not well developed.
To circumvent the difficulty, we
add small diffusion to both equations and consider the regularized model
\beq
\partial_{t}h=\frac{m\psi (|\nabla h|)}{\nu}+\varepsilon _{h}\Delta h,\label{hr}
\eeq
\beq
\lambda \partial _{t}m-\nabla \cdot (m\nabla h)=w-\frac{m\psi (|\nabla
h|)}{\nu }+\varepsilon _{m}\Delta m,  \label{mr}
\end{equation}
where the positive coefficients $\varepsilon_h(\nu)$ and $\varepsilon_m(\nu)$ vanish as 
$\nu$ tends to zero.
It should be noted that, although small diffusion may be physically meaningful
and has been included into the original BCRE formulation \cite{BCRE1,BCRE2}, here we
introduce it merely as a parabolic regularization of hyperbolic equations convenient for
analyzing the
model's behavior at $\nu\rightarrow 0$.

We assume the same initial and boundary conditions,  (\ref{t0}) and
(\ref{bound}) correspondingly, for the function
$h$. The non-negative values of $m$ both in $\Omega$ at  $t=0$ and on the boundary
of this domain for $t>0$ may be chosen
arbitrary: these initial and boundary conditions result only in
the appearance of boundary layers in the solution for any finite $\nu >0$ and are lost 
in the $\nu \rightarrow 0$ limit. 
Rigorously, convergence of the problem (\ref{hr})-(\ref{mr}) to the variational
inequality (\ref{vi}) is proved elsewhere \cite{tobe}. Here we present 
a simplified scheme of the
proof and avoid technicalities.

The main step
is, as usual, to obtain uniform in $\nu >0$ {\it a priori} estimates on the
solutions of the equations (\ref{hr})-(\ref{mr}). 
First, taking the gradient and multiplying by $\nabla h$, we derive from (\ref{hr}) a
parabolic partial differential equation for $|\nabla h|^2$. 
Since $\psi(k)=0$ and $|\nabla h_0|\le k$, we are able, using the  
maximum principle for this equation, to show that for $\nu>0$ 
\beq
|\nabla h|\le k\ \mbox{\it for all}\ (x,t);\ h\ \mbox{\it is
uniformly bounded.}\label{Max1}
\eeq
Second, using the non-negativeness of the source function $w(x,t)$ and
applying the maximum principle to the equation (\ref{mr}), we deduce that 
for each $\nu>0$
\begin{equation}
m\geq 0\text{\it \ for all }(x,t).  \label{Max2}
\end{equation}
Applying the estimates (\ref{Max1}), (\ref{Max2}) to the equation (14) we
obtain 
\begin{equation}
m\psi (|\nabla h|)=O(\nu ).  \label{mp}
\end{equation}
Sending {\it \ }$\nu $ to zero in (\ref{Max1}), (\ref{Max2}), (\ref{mp}) we
establish  the fulfillment in this limit of the conditions (\ref{m}), (\ref{cond1}), 
and
(\ref{cond2}).
Finally, adding the equations (\ref{hr}) and (\ref{mr}) we obtain
\[
\lambda \partial_{t}m+\partial_{t}h-\nabla \cdot (m\nabla h)=w+\varepsilon _{h}
\Delta h+\varepsilon _{m}\Delta m.
\]
Since $\lambda$, $\varepsilon _{h}$, and $\varepsilon _{m}$ vanish as $\nu \rightarrow
0$, we can show  that the corresponding limits of $h$ and  $m$ satisfy also the
balance equation (\ref{conserv}) in some weak (integral) sense. 
This completes the proof, because the model (\ref{m})-(\ref{bound})
is
equivalent to the variational inequality ({\ref{vi}).

To illustrate this result we will now compare 
solutions of the BCRE-type model (\ref{hr})-(\ref{mr}),
solutions of the variational inequality (\ref{vi}), and 
real shapes of small and large piles. Let us consider the simplest
situation: a pile growing under a point source on an infinite horizontal
support $h_0=0$. Although in this case the piles are known to be almost 
perfect cones, sometimes
one can notice \cite{AH} curved tails near the bottom of a small 
pile (Fig.~\ref{cone}a).
As the pile becomes larger, the tail remains  
of only, say, tens of grain diameters long, so
the tail of a large pile is difficult to see (Fig.~\ref{cone}b). 
\begin{figure}[htb]
\centerline{\psfig{figure=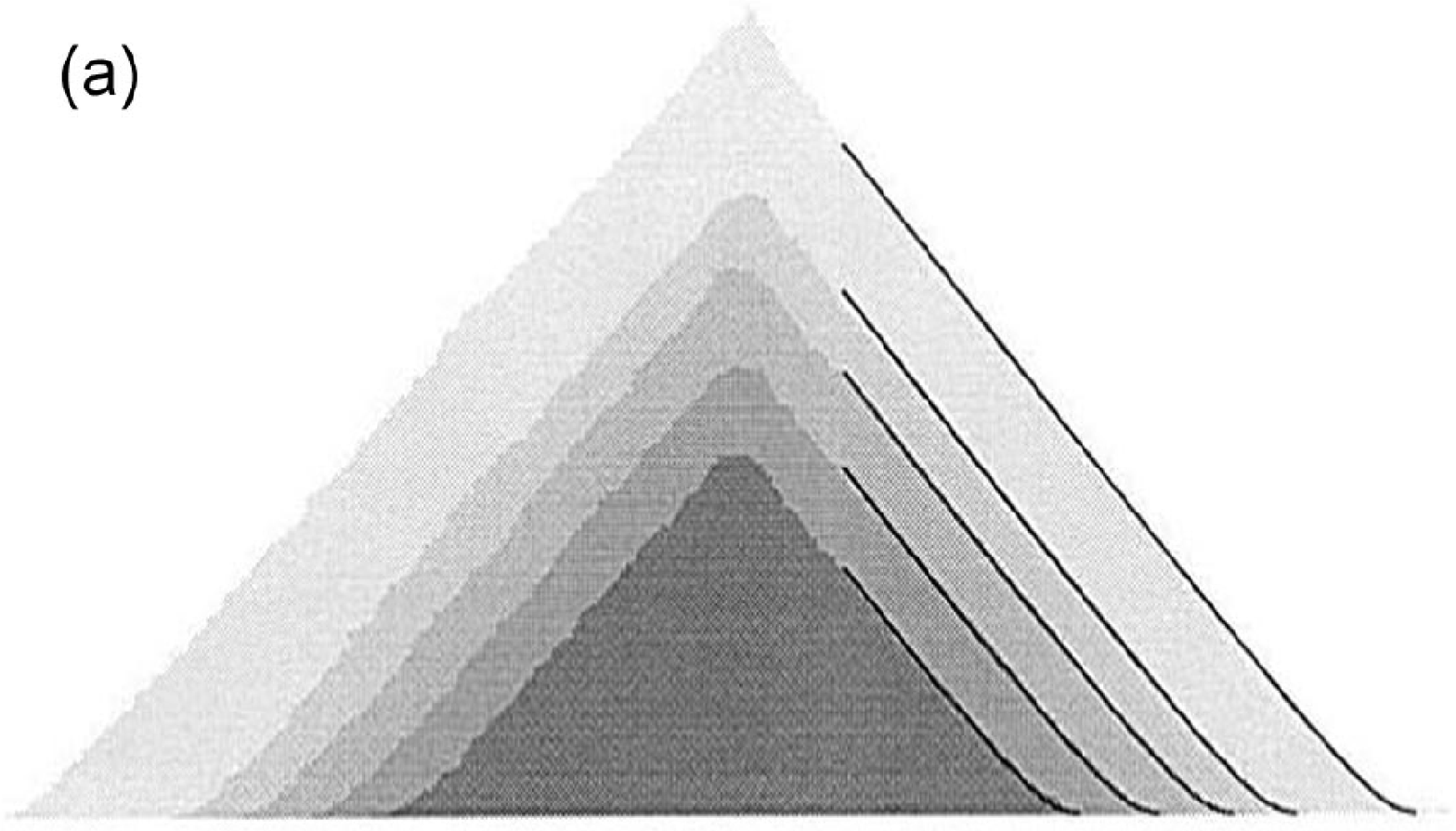,bbllx=35pt,bblly=236pt,bburx=577pt,bbury=556pt,width=8.4cm}}
\vspace{.2cm}
\centerline{\psfig{figure=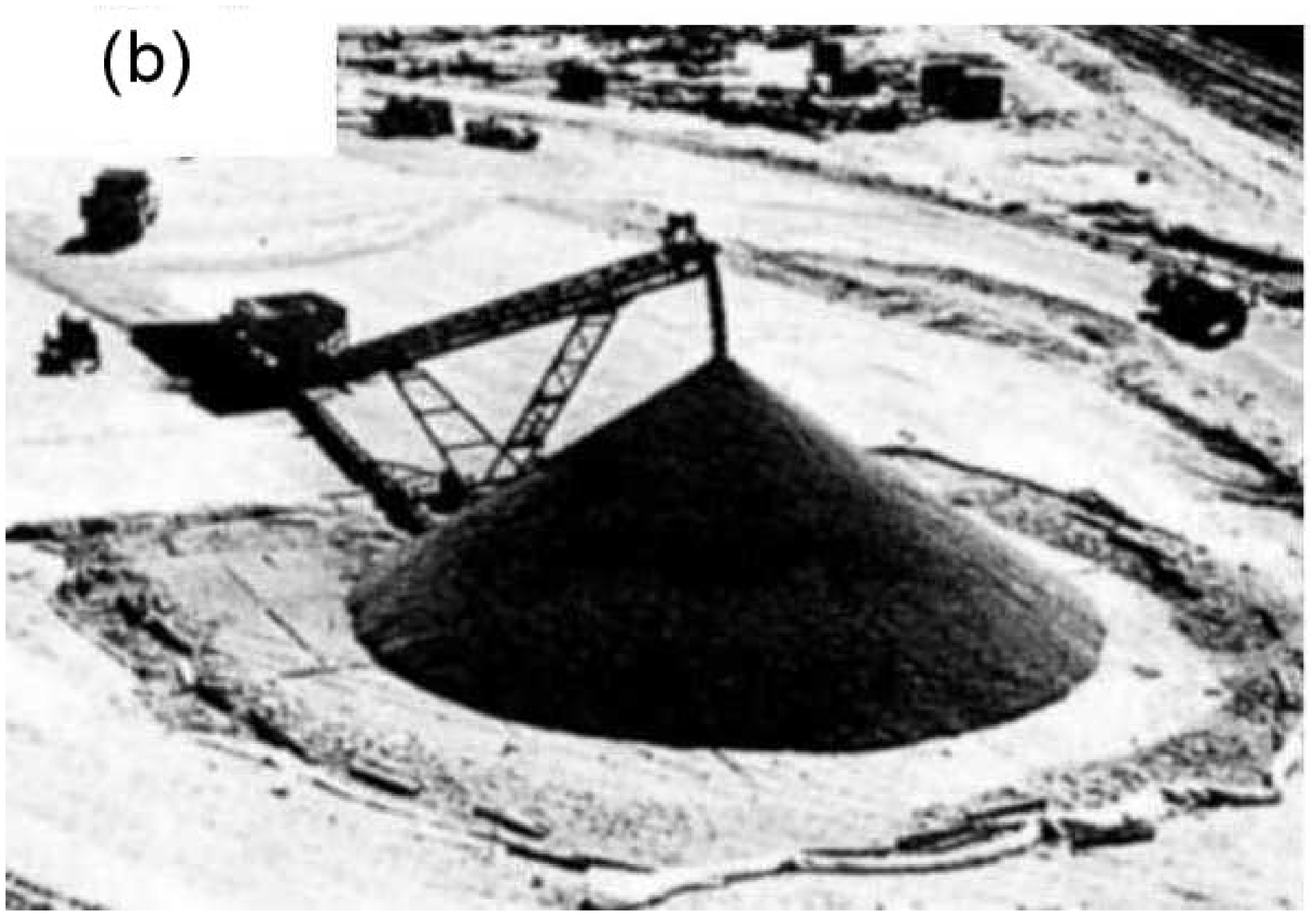,bbllx=35pt,bblly=208pt,bburx=572pt,bbury=584pt,width=8.4cm}}
\caption{Piles growing below a point-like source.
(a) Digitized image of a small polenta pile, Alonso and Herrmann [32]. Different gray 
levels show the pile at different stages of growth, the curved tails near 
the pile bottom remain short; (b) Large conical pile, photo by Slater [33].}
\label{cone}
\end{figure}
The modified BCRE model (\ref{hr})-(\ref{mr})
describes this situation quite satisfactory, see Fig. \ref{psource}. 
Although the tails of small piles ($\nu=0.2$) are
clearly
seen, tails of the larger piles ($\nu=0.01$) is difficult to detect.
We see also that
piles, predicted by the BCRE model with small $\nu$ and $\lambda$,
are very close to the growing cone,
exact analytical solution of the
variational inequality (\ref{vi}).
It may be noted that for small values of $\nu$ and $\lambda$ the model equations
are stiff and their numerical solution becomes difficult.
Thus, even using an implicit finite-difference approximation of (\ref{hr})-(\ref{mr}),
we needed $10^5$ time steps in the latter example. 
\begin{figure}[hbt]
\centerline{\includegraphics[width=8.5cm]{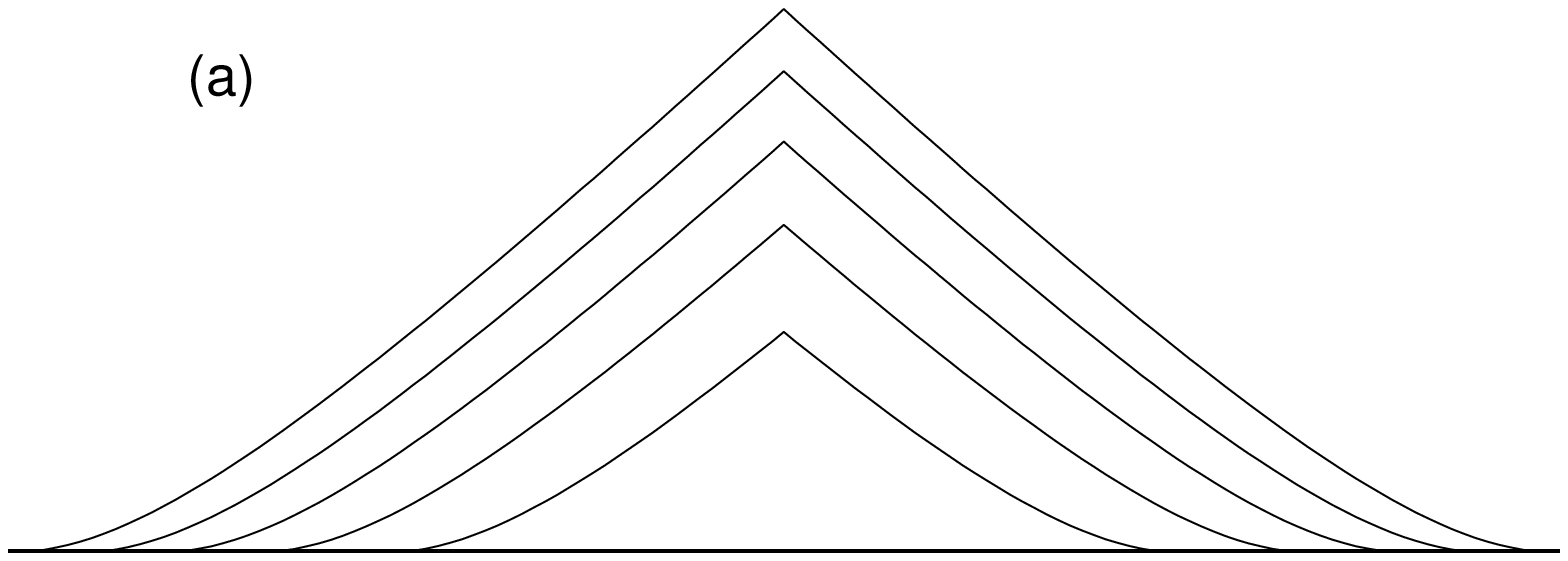}}
\vspace{.3cm}
\centerline{\includegraphics[width=8.5cm]{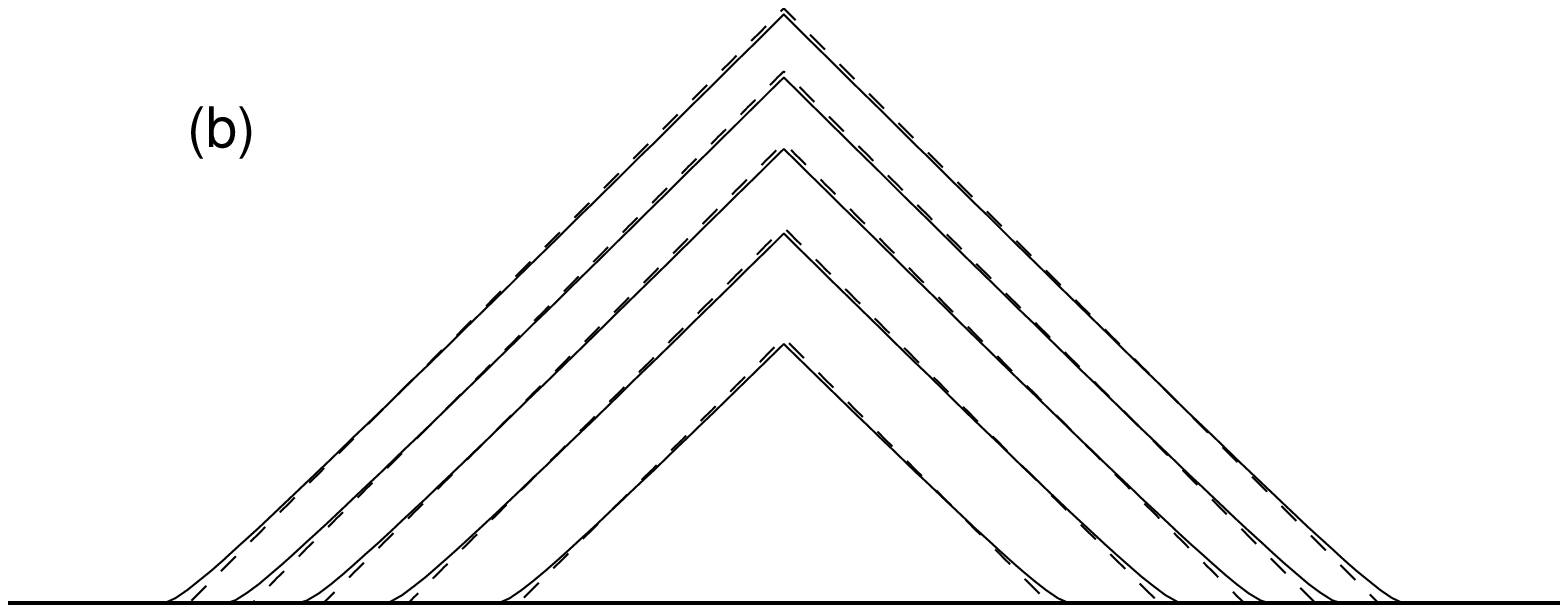}}
\vspace{.2cm}
\caption{Piles growing below a point-like source,
numerical solution of (\ref{hr})-(\ref{mr}) with
$\varepsilon_h=\varepsilon_m=0, \ m|_{t=0}=0, \ \lambda=0.1\nu$, and $k=1$.\\ (a) Small
piles, $\nu=0.2$; (b) Solid line -- large piles, $\nu=0.01$;
dashed line -- analytical solution of the variational
inequality (\ref{vi}).}
\label{psource}
\end{figure}

\section{Conclusion}
We considered two different continuous models
for the pile surface dynamics: the BCRE model \cite{BCRE1,BCRE2} and the variational model
\cite{ZhVM,PRE1}. Both models are written for two coupled 
dependent variables and are able, in principle, to
account for multiplicity of metastable pile shapes and surface
avalanches. It has been found that the models are
related and describe the
pile surface dynamics on different spatio-temporal scales. 

BCRE-type models
may be used to simulate the fast processes, such as the 
initiation, spreading, and settling down of an avalanche.
To describe the much slower dynamics of the mean shape of a pile, 
the model may often be simplified by employing a
quasistationary equation
for the rolling grains layer. Such a model is able to predict some peculiarities of small 
pile shapes \cite{AH} and was recently employed for simulating 
the nonlinear dynamics of sand ripples \cite{Ripple}.

Unlike the BCRE models, the variational model of pile growth does not permit the 
discharged grains to roll upon subcritical slopes and 
is therefore unable to account for such features of sand surface as sand-ripple instability
or surface slope deviation from the critical angle near the bottom of a conical pile. 
Indeed, these effects are determined by rolling of particles upon the
subcritical slopes and are exhibited
on the length scale comparable to the
mean path of a particle prior to its being trapped.

On the other hand, sand ripples on the  dune surface or tiny tails at the bottom of a 
pile are seen only from a short distance. These small details are 
difficult to distinguish watching from a
larger distance allowing one to follow the evolution of a big dune or formation
of a 
large pile. In such situations the BCRE model
contains another small parameter. This complicates simulations and makes them
inefficient.
As has been shown in our work, in the long-scale limit the BCRE-type 
models converge
to the variational model of pile growth. The latter model is more appropriate for 
simulating the pile surface dynamics on a large spatio-temporal scale.


\begin{thebibliography}{99} 

\bibitem{ZhVM} L. Prigozhin, {Zh. Vychisl. Mat. Mat. Fiz.}
{\bf26}, 
1072 (1986).
\bibitem{PRE1} L. Prigozhin, {Phys. Rev. E}, {\bf 49}, 1161 (1994).
\bibitem{Euro1} L. Prigozhin, {Euro. J. Appl. Math.} {\bf 7}, 225 (1996). 
\bibitem{ChEngSci} L. Prigozhin, {Chem. Eng. Sci.} {\bf 48}, 3647 
(1993).
\bibitem{EV} T. Elperin and A. Vikhansky, {Phys. Rev. E} {\bf 55},
5785 
(1997).
\bibitem{Puhl} H. Puhl, {Physica A} {\bf 182}, 295 (1992).
\bibitem{Smith} K. Smith, {\it Environmental Hazards} (Routledge, 
London -- New York, 1996).
\bibitem{Evans1} G. Aronsson, L.C. Evans, and Y. Wu, {J. Differ. 
Equations} {\bf 131}, 304 (1996).
\bibitem{Evans2} L.C. Evans, M. Feldman, and R.F. Gariepy, {J. 
Differ. Equations} {\bf 137}, 166 (1997).
\bibitem{Evans3} L.C. Evans and F. Rezakhanlou, {Commun. Math. 
Phys.} {\bf 197}, 325 (1998).
\bibitem{Bak} P. Bak, C. Tang, and K. Wisenfeld,
{Phys. Rev. Lett.} {\bf 59}, 381 (1987).
\bibitem{BCRE1} J.-P. Bouchaud, M.E. Cates, J. Ravi Prakash, and 
S.F. Edwards,  {J. Phys. I France} {\bf 4}, 1383 (1994).
\bibitem{BCRE2} J.-P. Bouchaud, M.E. Cates, J. Ravi Prakash, and
S.F. Edwards,  {Phys.
Rev. Lett.} {\bf 74}, 1982 (1995).
\bibitem{deG1} P.-G. de Gennes, C.R. Acad. Sci., Ser. II-B, {\bf 321}, 501 (1995).
\bibitem{deG2} T. Boutreux and P.-G. de Gennes, C.R. Acad. Sci., Ser. II-B, {\bf 325},
85 (1997);
\bibitem{deGA}P.-G. de Gennes, Avalanches of granular materials, in: 
R. Behringer and J.T. Jenkins (eds), {\it Powders and Grains 97} 
(Balkema, Rotterdam, 1997).
\bibitem{Thick} T. Boutreux, E. Raphael, and P.-G. de Gennes, 
{Phys. Rev. E} {\bf 58}, 4692 (1998); 
P.-G. de Gennes, Physica A {\bf 261}, 267 (1998);
A. Aradian and P.-G. de Gennes,
{Phys. Rev. E} {\bf 60}, 2009 (1999).
\bibitem{MP} L. Mahadevan and Y. Pomeau, {Europhys. Letters} {\bf 
46}, 595 (1999);
T. Emig, P. Claudine, and J.-P. Bouchaud, {Europhys. Letters}, to
appear.
\bibitem{BC}  J.-P. Bouchaud and M.E. Cates, Granular Matter {\bf 1},
101 (1998).
\bibitem{DD} A. Daerr and S. Douady, Nature (London) {\bf 399}, 241 (1999).
\bibitem{JN} H.M. Jaeger and S.R. Nagel, Science {\bf 255}, 1523 (1992).
\bibitem{DL} G. Duvout and J.-L. Lions, {\it Les In\'{e}quations en M\'{e}canique et
en
Physique} (Dunod, Paris, 1972).
\bibitem{GLT} R. Glowinski, J.-L. Lions, and R. Tr\'{e}moli\`{e}rs,
{\it Analyse Num\'{e}rique des In\'{e}quations 
Variationnelles} (Dunod, Paris, 1976).
\bibitem{FBP} L. Prigozhin, Free Boundary Problems News, no. 10, 
2 (1996).
\bibitem{DM} S.N. Dorogovtsev and J.F.F. Mendes, Phys. Rev. Lett.
{\bf 83}, 2946 (1999).
\bibitem{Ripple} L. Prigozhin, Phys. Rev. E {\bf 60}, 729 
(1999).
\bibitem{HK} K.P. Hadeler and C. Kuttler, Granular Matter {\bf 2},
9 (1999).
\bibitem{rough}F.-X. Riguidel, R. Jullien, G.H. Ristow, A. Hansen, and D. Bideau,
J. Phys. I France {\bf 4}, 261 (1994);
S. Dippel, G.G. Batrouni, and D.E. Wolf, Phys. Rev. E {\bf 56}, 3645 (1997);
 C. Henrique {\it et al.}, Phys. Rev. E
{\bf 57}, 4743 (1998). 
\bibitem{AI} M.A. Aguirre {\it et al.}, Powder Technology {\bf 92}, 75 (1997).
\bibitem{P} O. Pouliquen, Physics of Fluids {\bf 11}, 542 (1999).
\bibitem{tobe} L. Prigozhin and B. Zaltzman, in preparation.
\bibitem{AH} J.J. Alonso and H.J. Herrmann, Phys. Rev. Lett. {\bf 76}, 4911 (1996).
\bibitem{Slater} R.A. Slater, Stockpiling and Reclaiming Systems, 
in: 
M.C. Hawk (ed.), {\it 
Bulk Materials Handling}, v. 1 (University of Pittsbourhg, Pittsbourhg, 1971).
\end{thebibliography}
\end{document}